\documentclass[prl,twocolumn]{revtex4}
\usepackage{hyperref}
\usepackage{amssymb}
\usepackage{graphicx}
\usepackage{color}

\newcommand{\avg}[1]{\left\langle {#1} \right\rangle}

\newcommand{\beq}{\begin{equation}}
\newcommand{\eeq}{\end{equation}}

\begin{document}

\title{Advantages of Coherent Feedback for Cooling Quantum Oscillators}

\author{Ryan Hamerly}\email{rhamerly@stanford.edu}
\author{Hideo Mabuchi}\email{hmabuchi@stanford.edu}
\affiliation{Edward L.\ Ginzton Laboratory, Stanford University, Stanford, CA 94305}
	
\date{\today}
	
\begin{abstract}
We model the cooling of open optical and optomechanical resonators via optical feedback in the Linear Quadratic Gaussian setting of stochastic control theory. We show that coherent feedback control schemes, in which the resonator is embedded in an interferometer to achieve all-optical feedback, can outperform the best possible measurement-based schemes in the quantum regime of low steady-state excitation number. Such performance gains are attributed to the coherent controller's ability to process non-commuting output field quadratures simultaneously without loss of fidelity, and may provide important clues for the design of coherent feedback schemes for more general problems of nonlinear and robust control.
\end{abstract}

\maketitle	

Feedback control of classical dynamical systems plays a central role in modern engineering~\cite{AstromMurray} but its quantum analogue, the notion of controlling a quantum system via feedback with a quantum or classical controller, is much less developed. Recent progress in modeling~\cite{Bela83,Wise93,Dohe00,WM09} and realizing~\cite{Smit02,Arme02,Bush06,Giga06,Arci06,Klec06,Mabu08,Gill10,Sayr11,Iida11} quantum feedback underscores the need for systematic approaches to control design, as do the wide range of potential applications in quantum science and technology.

While some of the most interesting problems in quantum feedback control are nonlinear ~\cite{BitFlipPaper,BaconShorPaper,Mabu11a,Goug11}, linear open quantum systems provide a logical first step towards more general problems. Working with linear systems, James, Nurdin and Petersen~\cite{JamesNurdinPetersen08,NurdinLQG} have utilized interconnection models based on quantum stochastic differential equations (QSDEs)~\cite{Huds84,Carm93,Gard93,Barc06} to develop generalizations of the traditional ${\cal H}^\infty$ and Linear Quadratic Gaussian (LQG) control paradigms that allow for the possibility of coherent optical feedback with linear quantum controllers. Here we work within the quantum LQG framework of~\cite{NurdinLQG} to study steady-state cooling of open quantum oscillators such as optical and optomechanical resonators subject to stationary heating and damping, with optical probing and feedback. We utilize numerical optimization together with fundamental analytic results~\cite{AstromMurray, Simon06} to describe coherent feedback control schemes that outperform the best possible measurement-based schemes. We find more systematic and quantitatively significant advantages of coherent feedback over measurement-based feedback than in linear control scenarios considered previously~\cite{Mabu08,NurdinLQG}, and interpret these performance gains in terms of the way that non-commuting field quadratures propagate through the feedback loop.


\begin{figure}[b]
	\centering
	\includegraphics[width=1.00\columnwidth]{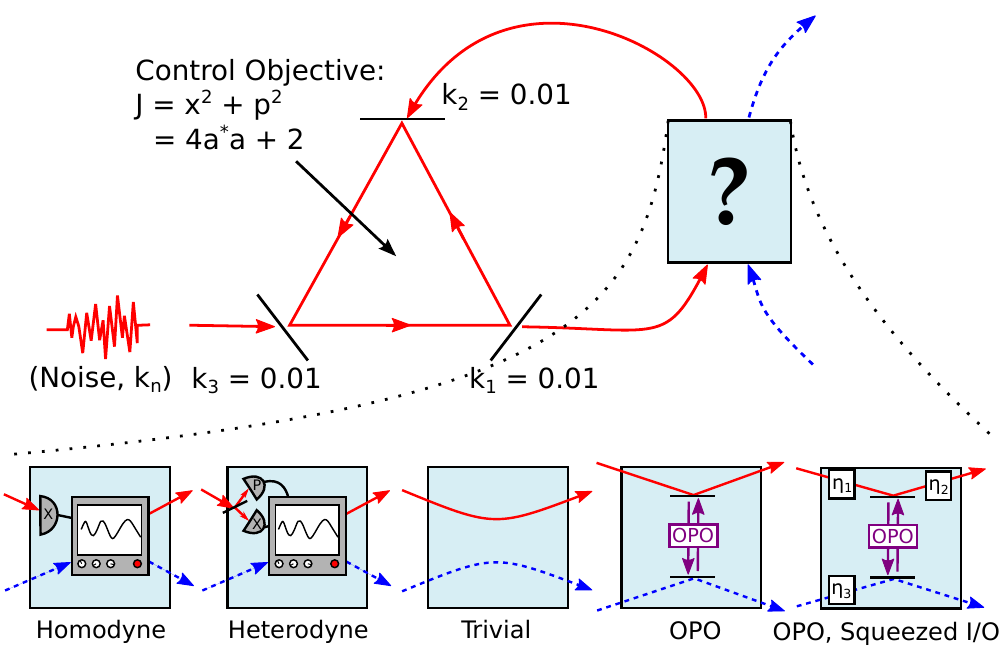}
	\caption{Possible Classical and Coherent Cavity Controllers for an Optical Cavity Plant System}
	\label{fig:f1b}
\end{figure}

We model quantum harmonic oscillators as cascadable open quantum systems using the SLH framework \cite{Gough08,Gough09} and associated QSDEs. In the SLH framework, any open quantum system may be described as a triple $G = (S, L, H)$ where $S$ is a scattering matrix, $L$ is a coupling vector and $H$ is the Hamiltonian operator for the system's internal degrees of freedom.  For a {\em linear} system with a vector of internal state variables $x$, the quantum stochastic differential equations (analogues to the Heisenberg Equations for open quantum systems \cite{Huds84, Gardiner92}) take the following form~\cite{JamesNurdinPetersen08,NurdinLQG}:
\begin{eqnarray}
	dx(t) & = & \left[A\,x(t) + a\right] dt + B\,da(t) \nonumber \\
	d\tilde{a}(t) & = & \left[C\,x(t) + c\right] dt + D\,da(t) \label{eq:abcd}
\end{eqnarray}
Here $A$, $B$, $C$, $D$, $a$ and $c$ are real and related to $(S, L, H)$.  The processes $da(t)$ and $d\tilde{a}(t)$ are Hermitian quantum stochastic processes for the inputs and outputs, respectively, defined by $da_i = \bigl(dA_i + dA_i^\dagger, (dA_i - dA_i^\dagger)/i\bigr)$, where $dA(t)$ is the quantum Wiener process~\cite{Gard04,Bout07} following the It\^{o} rule $dA_i\,dA_j^\dagger = \delta_{ij} dt$ for vacuum inputs.

As a first example of a quantum control system, take a noisy optical cavity, Fig.\ \ref{fig:f1b}.  This has the SLH model
\beq
	S = I,\ \ L = \left[\sqrt{k_1} a,\ \sqrt{k_2} a,\ \sqrt{k_3} a\right],\ \
	 H = \Delta a^\dagger a,
\eeq
where $\Delta$ is the detuning of the cavity resonance frequency from that of a rotating frame. In the controller's absence, the cavity is driven by two vacuum inputs (mirrors $k_1$ and $k_2$) and one thermal input (mirror $k_3$), and the cavity relaxes to a thermal state where the mean photon number $\avg{a^\dagger a}$ depends on the noise power $k_n$.  The objective in this control problem is to minimize the effect of the noise on the cavity's internal state -- in other words, to minimize the photon number $\avg{a^\dagger a}$ of the cavity.  We accomplish this by sending output $1$ through a control circuit and feeding the result back into input $2$.  This is an LQG feedback control problem, assuming negligible control cost.

Once the full system is set up, with its associated $A$, $B$, $C$ and $D$ matrices, the covariance matrix $\sigma_{ij} = \frac{1}{2} \avg{x_i x_j + x_j x_i}$ can be computed with the Lyapunov equation, $A \sigma + \sigma A^{\rm T} + B F B^{\rm T} = 0$, where $F$ is the input noise covariance; the mean photon number can then be computed from $\sigma$.  Without a controller present, we find:
\beq
	\avg{a^\dagger a}_{\rm nc} = \frac{k_3}{k_1 + k_2 + k_3} k_n \label{eq:cav-nc}
\eeq

Five possible controllers are sketched in Figure \ref{fig:f1b}.  The classical controllers (Homodyne and Heterodyne) work by measuring a quadrature from the cavity's output, or splitting the beam and measuring two different quadratures, and applying a feedback signal based on this measurement and the controller's internal state.  The ``trivial controller'' works by feeding the output directly back into mirror 2 of the plant, perhaps with a phase shift.  The remaining two controllers function as coherent controllers with memory, where the feedback signal depends not only on the probe field, but also on its history.  The OPO with squeezed input/output (I/O) is the most general system realizable for two inputs and one internal degree of freedom~\cite{NurdinNetworkSynthesis}.

\begin{figure}[t]
	\centering
	\includegraphics[width=1.00\columnwidth]{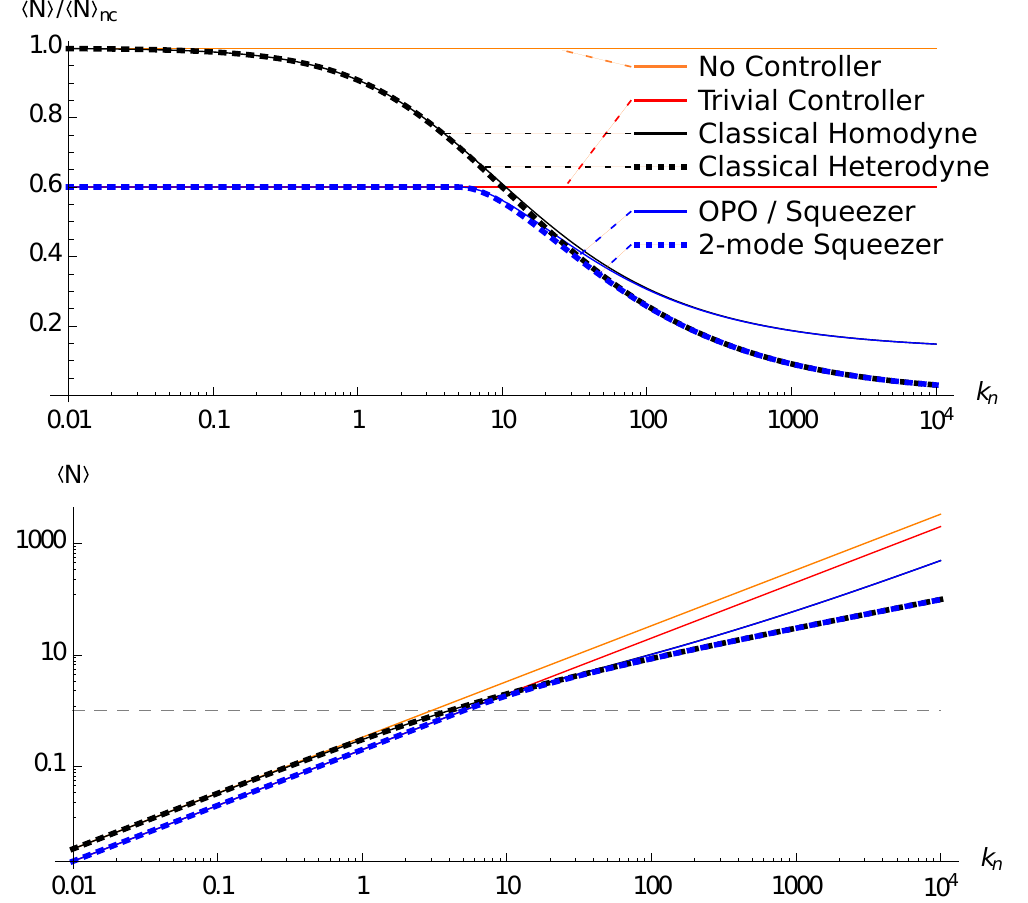}
	\caption{Bottom: Cavity photon number as a function of noise strength $k_N$, for various control schemes.  Top: Photon number relative to the no-control case.  Smaller is better.}
	\label{fig:f4}
\end{figure}

The classical LQG optimization problem is convex, and the optimal controller can be expressed analytically in terms of the plant parameters and the LQR~\cite{Simon06}. The optimal homodyne and heterodyne controllers act as classical amplifiers.  In the heterodyne case, the output $d\tilde{a}$ is related to the input $da$ and vacuum noises $da_{k1}$, $da_{k2}$ by: 
\begin{eqnarray}
	d\tilde{a}_x & = & \xi (da_x + da_{k1,x}) + da_{k2,x} \nonumber \\
	d\tilde{a}_p & = & \xi (da_p - da_{k1,p}) + da_{k2,p}
\end{eqnarray}
In other words, the controller measures the probe, amplifies it, and then adds the signal to a vacuum channel.  With this controller, the plant mean photon number is
\beq
	\avg{a^\dagger a}_{\rm cl} = \frac{k_2 \sinh^2\eta + k_3 k_n}{k_1 + k_2 + k_3 + 2\sqrt{k_1 k_2} \sinh\eta} \label{eq:lqr-cav-cl}
\eeq
which is plotted in dotted black in Figure \ref{fig:f4}.  Note that, while the controller does a good job in the classical regime, when $\avg{a^\dagger a}$ is high, it is ineffective in the quantum regime when the photon number is $\lesssim 1$.

Contrast this with the ``trivial controller'', which feeds the output from mirror $1$ directly into mirror $2$.  Rather than leaking photons separately, the two outputs then interfere constructively~\cite{Mabu11a}.  This leads to an enhancement in the dissipation and a mean photon number
\beq
	\avg{a^\dagger a}_{\rm tr} = \frac{k_3}{k_1 + k_2 + k_3 + 2\sqrt{k_1 k_2}} k_n \label{eq:cav-tr}
\eeq
which beats the best classical controller by a constant factor.

A more sophisticated control design, the OPO with squeezed I/O, can be optimized numerically.  We find a (locally) optimal solution which is equivalent to a two-mode squeezer:
\begin{eqnarray}
	d\tilde{a}_{1} & = & \cosh(\eta) da_1 + \sinh(\eta) da_2 \nonumber \\
	d\tilde{a}_{2} & = & \sinh(\eta) da_2 + \cosh(\eta) da_2
\end{eqnarray}
This gives a mean photon number of:
\beq
	\avg{a^\dagger a}_{\rm 2-sq} = \frac{k_2 \sinh^2\eta + k_3 k_n}{k_1 + k_2 + k_3 + 2\sqrt{k_1 k_2}\cosh\eta} \label{eq:sq-2md}
\eeq
which is also plotted in Figure \ref{fig:f4}.

Notice just how similar Equations (\ref{eq:lqr-cav-cl}) and (\ref{eq:sq-2md}) are.  Both the heterodyne controller and the linear amplifier reduce the cavity's photon number by amplifying the feedback signal, but also add noise to the system.  For equivalent levels of amplification, the classical controller adds extra noise into the system from the measurement process.  When $k_n$ and $\eta$ are large, this extra noise is negligible, but in the quantum regime where $k_n$ and $\eta$ are $\lesssim 1$, this noise can play a major role in making the linear amplifier outperform the heterodyne controller.

Of course, this need not be the global optimum.  It is possible that other solutions, not found in our search of the parameter space, lead to better LQG performance than the two-mode squeezer.  Nevertheless, it is significant that we have found a quantum controller which does better than the optimal classical controller for all values of $k_n$, particularly in the quantum regime where the excitation number is low.

\begin{figure}[b]
	\centering
	\includegraphics[width=1.00\columnwidth]{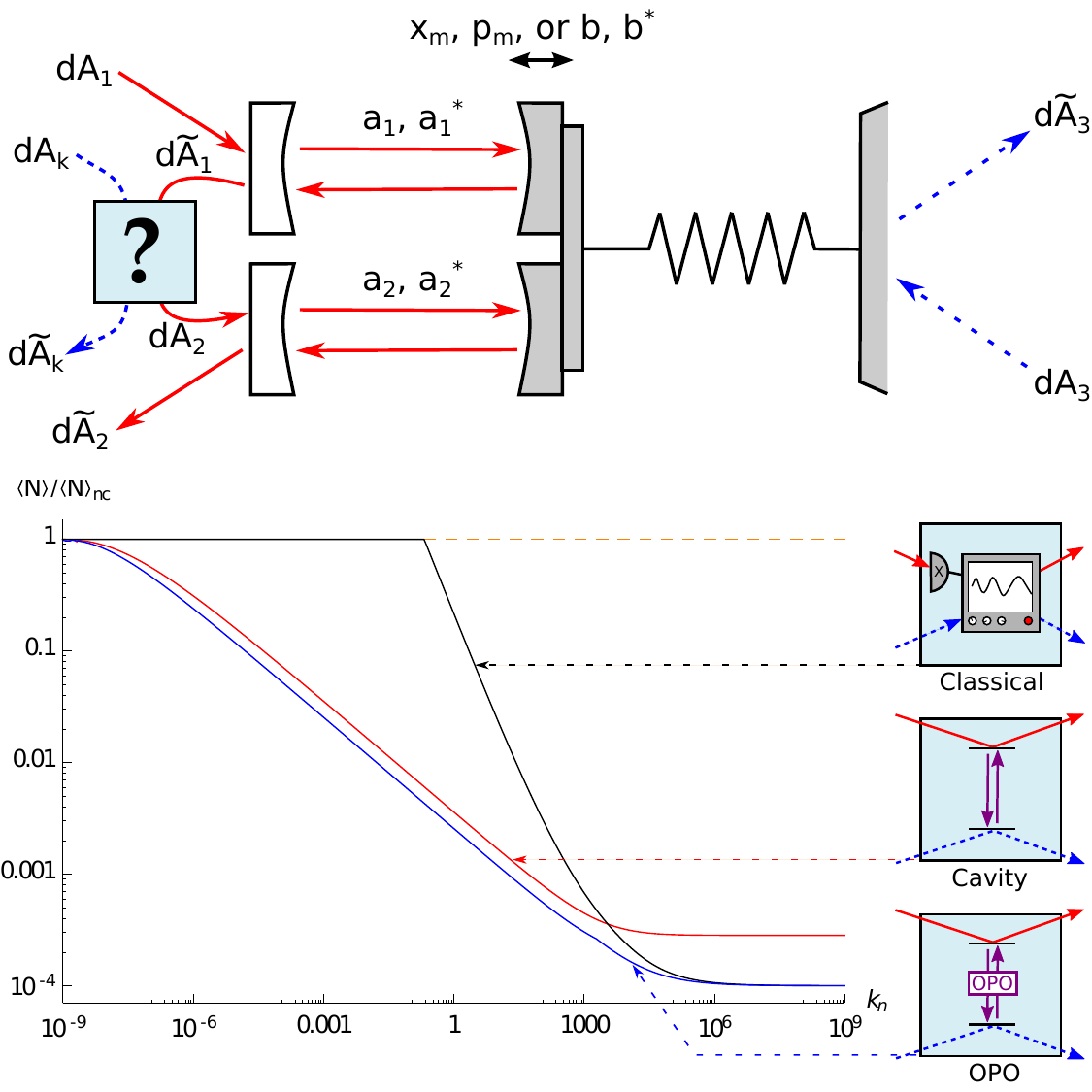}
	\caption{Control-system setup for the mechanical oscillator cooling problem with $\Omega = 100,\ Q = 10000$.  Three potential controller designs and their relative performance.}
	\label{fig:f17a}
\end{figure}

Another model system to consider is the optomechanical oscillator.  Optomechanical oscillators -- mechanical springs that couple to an optical field via a cavity -- have been a topic of tremendous recent interest in the physics community~\cite{MarquardtGirvin,Bott12}. A central goal has been to find ways to exploit optomechanical coupling to cool the mechanical oscillator from ambient temperature to its ground state.

Consider a single cavity coupled to a mirror on a spring.  If we shine a laser into the input port, and adiabatically eliminate the internal mode, we obtain the SLH model
\beq
	S = 1_{2\times 2},\ \ \ L = \left[K x_m, \sqrt{\Omega/Q}b\right],\ \ \	H = \Omega b^\dagger b \label{eq:sand-cav}
\eeq
where $K$ is a constant depending on the cavity and the drive field.  It is not hard to show that this gives the following input-output relations:
\begin{eqnarray}
	& & \left\{\begin{array}{rcl}
		dx_m & = & \Omega p_m - \Omega/Q x_m \\
		dp_m & = & -\Omega x_m - \Omega/Q p_m - 2K da_p
		\end{array}\right. \nonumber \\
	& & \left\{\begin{array}{rcl}
		d\tilde{a}_{x} & = & da_{x} + 2K x_m dt \\
		d\tilde{a}_{p} & = & da_{p}
		\end{array}\right.
\end{eqnarray}
The state variable $x_m$ is imprinted on the output $d\tilde{a}_{x}$, permitting the controller to be used as a ``measurement'' device via the $x$ quadrature.  Conversely, the $p$-quadrature input $da_{p}$ alters the state of the mirror, allowing the system to function as a ``feedback'' device.

Coupling two such cavities together as in Fig.~\ref{fig:f17a}, we obtain our model plant system.  The cooling objective is to minimize the phonon number $\avg{b^\dagger b}$, where $b$ is the spring's annihilation operator.  As in the optical cavity case, this is an LQG control problem.

\begin{figure}[b]
	\centering
	\includegraphics[width=1.00\columnwidth]{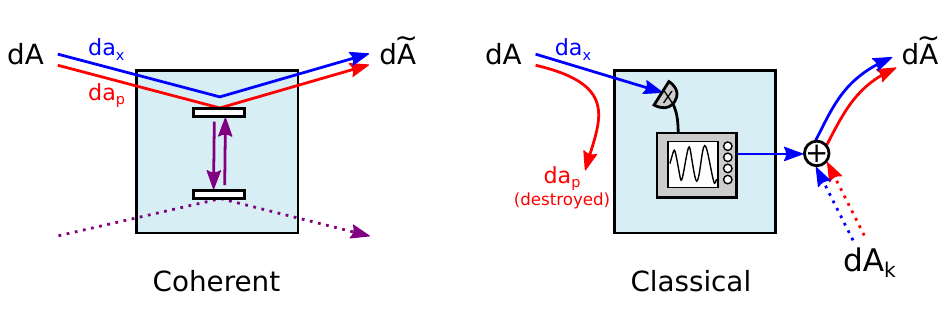}
	\caption{Flow of $x$- and $p$-quadrature signals (blue and red, respectively) in the classical and coherent controllers.}
	\label{fig:f16}
\end{figure}

Consider first the classical controller, black in Fig.~\ref{fig:f17a}.  This controller only measures the $da_x$ quadrature, since no information is contained in $da_p$, and applies a feedback based on this input (such strategies outperform heterodyne-based classical control). When optimizing, we vary both the controller parameters {\it and} the couplings $K_1$, $K_2$ to the cavities in (\ref{eq:sand-cav}), since these depend on input laser powers rather than on properties of the plant. Here we will assume $K_1 = -K_2 \equiv K$, however, as this type of design could be implemented practically using two modes of a single optical cavity with comparable power circulating in each. This symmetry in the control model also facilitates straightforward interpretation of the advantage of coherent feedback.

As before, the classical controller does a good job cooling the oscillator in the classical limit $\avg{N} \gg 1$, but does a very poor job in the quantum limit; below a threshold value, it has no effect at all.  The classical controller adds two sources of noise to the plant.  First, the probe field adds {\it measurement noise} to the spring through the first cavity.  Second, the feedback field adds a {\it feedback noise} of equal magnitude via the second cavity.  These noises add up incoherently.  In the low-$k_n$ limit, they dominate the dynamics of the control system and the classical controller becomes ineffective.

We also optimized two types of coherent controller -- an optical cavity and an OPO.  They both behave qualitatively in a similar way, but the optical cavity is easier to understand. With the optical cavity controller, the feedback field is related to both the probe field and the cavity state, which changes with time:
\begin{eqnarray}
	da & = & (-i\Delta - \kappa/2)a\,dt + \sqrt{\kappa} d\tilde{A}_1 \nonumber \\
	dA_2 & = & d\tilde{A}_1 + \sqrt{\kappa}a\,dt
\end{eqnarray}
Optimization shows that $\Delta=\Omega$, so both the cavity and mirror resonate at the same frequency. Although not interpreted in terms of feedback control theory, it is straightforward to show that cavity-optomechanical cooling experiments such as~\cite{Giga06,Arci06,Klec06} have implemented an equivalent strategy in which the three optical cavities of Fig.~\ref{fig:f17a} are collapsed into one.

The LQG-optimal classical controller relies on a Kalman filter that recursively estimates the plant state from the measured signal~\cite{AstromMurray}. The coherent cavity controller can also be thought of as a Kalman filter, but one that preserves the coherence of the input signal $d\tilde{A}_1$. In the classical controller the $p$-quadrature $da_p$ is essentially discarded after the measurement, while in the cavity controller the field retains its coherent properties and $da_p$ coming out is the same as $da_p$ going in. This correlates the noises in the measurement and feedback cavities such that the associated forces on the mechanical oscillator exactly cancel, as shown in Fig.~\ref{fig:f16}. This cancellation of the measurement noise is what gives the coherent cavity controller its superior performance in the low phonon-number regime.

LQG control is not the first problem to benefit from this noise cancelation.  Previous studies had shown that precision measurement schemes~\cite{TsengCaves}, particularly in the context of LIGO~\cite{Arcizet06}, show similar improvements.

The OPO controller is more general than the cavity controller and its performance is slightly better over the whole noise range.  As with the optical cavity case, it is significant that we have found realistic coherent control schemes that perform significantly better than provably optimal measurement-based schemes.

In this paper, we have studied the coherent control of linear quantum systems from an LQG perspective.  In the systems studied, we have shown that there is always a quantum controller that does at least as well as the optimal classical controller. In the quantum regime, when excitation number in the plant is of order unity, we have shown that the best quantum controller can do better -- in some cases, significantly so.  One could straightforwardly extend these results to non-quadratic cost functions in linear control systems. Indeed, some work has already been done on this matter, focusing on using coherent feedback to maximize the squeezing in a cavity mode \cite{Iida11}. Viewing an optomechanical device as a control system may also provide insight into minimizing the noise in optomechanical sensors. Finally, the fact that coherent controllers perform better because they utilize both quadratures of the input field may help guide the design of quantum controllers for nonlinear systems such as optical switches or error correcting codes.

\begin{acknowledgements}
This work has been supported by the NSF (PHY-1005386), AFOSR (FA9550-11-1-0238) and DARPA-MTO (N66001-11-1-4106). Ryan Hamerly is supported by the NSF GRFP and a Stanford Graduate Fellowship. We thank Nikolas Tezak, Gopal Sarma, Dmitri Pavlichin and Orion Crisafulli for useful discussions.
\end{acknowledgements}

\end{document}